\documentclass[aps,prb,twocolumn,superscriptaddress]{revtex4-1}

\usepackage{hyperref} 
\usepackage{graphicx}
\DeclareGraphicsExtensions{.png,.jpg,.eps}
\usepackage{xcolor}

\begin{document}

\title{ On the origin of magnetic anisotropy in two dimensional CrI$_3$}

\author{J. L. Lado}
\affiliation{QuantaLab, International Iberian Nanotechnology Laboratory (INL),
Av. Mestre Jos\'e Veiga, 4715-330 Braga, Portugal}

\author{J. Fern\'andez-Rossier}
\affiliation{QuantaLab, International Iberian Nanotechnology Laboratory (INL),
Av. Mestre Jos\'e Veiga, 4715-330 Braga, Portugal}

\affiliation{ Departamento de F\'isica Aplicada, Universidad de Alicante, 03690 Spain}

\date{\today} 

\begin{abstract} 

The observation of ferromagnetic order in a monolayer of CrI$_3$ has been
recently reported, with a Curie temperature of 45 Kelvin and off-plane easy
axis.    Here we study the origin of magnetic anisotropy, a necessary
ingredient to have  magnetic order in two dimensions, combining 
two levels of modeling,   density functional calculations 
and spin model Hamiltonians.   We find two different contributions to the
magnetic anisotropy of the material, favoring off-plane magnetization and
opening a gap in the spin wave spectrum. First,  ferromagnetic
super-exchange  across the  $\simeq $ 90 degree Cr-I-Cr bonds, are anisotropic,
due to the spin orbit interaction of the ligand I atoms. 
Second,   a much smaller contribution  that comes from the single  ion anisotropy of the $S=3/2$  Cr atom.
Our results permit to establish the XXZ Hamiltonian,  with a very small single
ion anisotropy, as the adequate  spin model for this system. 
Using spin wave theory we estimate the Curie temperature and we highlight the
essential role  played
by the  gap that magnetic anisotropy induces on the magnon spectrum.

\end{abstract}
\maketitle


\section{Introduction}

The recent reports of ferromagnetic order in two different
two dimensional 
crystals,\cite{gong2017,huang2017} Cr$_2$Ge$_2$Te$_6$ and  CrI$_3$, together
with the report of
antiferromagnetic order\cite{FePS32016raman,lee2016} in FePS$_3$ a few months earlier,  mark the beginning
of a new chapter in the remarkable field of two dimensional materials.   These
discoveries extend significantly the list of electronically ordered two
dimensional crystals, that included already superconductors,\cite{lu2015,Ugeda2016}  charge
density waves materials\cite{xi2015} and ferroelectrics.\cite{chang2016}
In  addition,  there is an increasing amount
of computational studies  predicting magnetic order in large variety of two
dimensional materials, such as VS$_2$ and VSe$_2$, \cite{ma2012}
K$_2$CuF$_4$,\cite{sachs2013} and the family of MPX$_3$, with  M the 3d transition
metals and X a group VI atom.\cite{chittari2016} 
The integration of magnetically ordered 2D
crystals in Van der Waals heterostructures\cite{geim2013} opens a vast field of possibilities
for new physical phenomena and new device concepts, and is already starting to
be explored experimentally.\cite{zhong2017} 

Mermin and Wagner demonstrated  the absence of long range magnetic order in spin-rotational
invariant systems with short range exchange interactions.\cite{mermin66}
 Therefore, the observation of long range magnetic
order in two dimensional insulating materials  stresses the importance of a quantitative microscopic
understanding of magnetic anisotropy in these systems.  The breaking of spin
rotational invariance  can be due to three mechanisms,  dipolar
interactions,  single ion anisotropy and anisotropy of the
exchange interactions.     In the case of very strong single ion
anisotropy,   a description in terms of the Ising model could be possible,
which automatically entails a magnetically ordered phase  phase at finite
temperature, as  predicted by Onsager in his remarkable  paper.\cite{onsager44}
However, large single ion anisotropies are normally associated to partially
unquenched orbital moment of the magnetic ion, which only happens for specific
oxidation states and low symmetry  crystal environments, most notably in
surfaces\cite{rau2014reaching} or for rare earth atoms.\cite{jensen1991rare}

CrI$_3$ is a layered transition metal compound known to order
ferromagnetically, in bulk, at $T_c=61$ Kelvin.\cite{dillon1965,mcguire2015}
Ferromagnetic order has been shown to persist in mechanically exfoliated
monolayers of  CrI$_3$, with a Curie Temperature of $T_c=45$ Kelvin, as
determined by  magneto-optical measurements.\cite{huang2017}
In this work we model magnetic anisotropy in a monolayer of CrI$_3$. Since
dipolar interactions favor in-plane anisotropy, 
we focus on the study of
both single ion anisotropy and exchange anisotropies.  To do that,  we first
model the system with relativistic all electron density functional theory (DFT)
calculations  that include spin orbit interactions, essential to account for
magnetic anisotropy.  Our calculations permit to build an effective spin model
with three energy scales, 
the isotropic and anisotropic Cr-Cr exchange  couplings, $J$
the anisotropic exchange $\lambda$,
and the single ion anisotropy $D$. 
As we show below,  $J$ and $\lambda$ 
are non zero, whereas the single ion anisotropy $D$ is negligible.


\begin{figure*}
 \centering
                \includegraphics[width=0.9\textwidth]{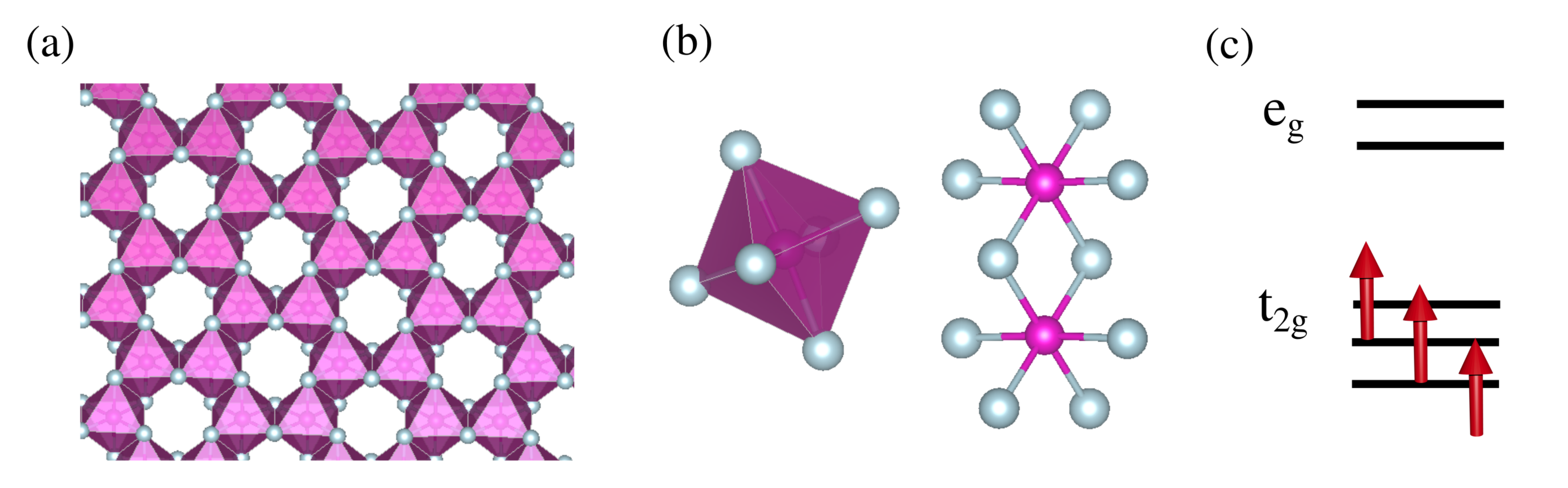}

\caption{(a) Crystalline structure of CrI$_3$, showing the
honeycomb arrangement of the chromium atoms. Every chromium atom
has an octahedral iodine environment, where different octahedra
are linked by an iodine forming an angle close to 90 degrees (b). 
The octahedral environment splits the d levels in the eg and t$_{2g}$ manifolds
(c).  First Hund rule favors the $S=3/2$ state, with 3 fully polarized electrons in the $t_{2g}$ manifold. 
}
\label{fig1}
\end{figure*}

Both experimental results \cite{mcguire2015,huang2017} and DFT
calculations\cite{zhang2015,wang2016doping}
show that CrI$_3$ is an semiconducting material
with a the band-gap of $1.2$ eV.\cite{dillon1965}  
In a single layer of CrI$_3$,  the plane of Cr
atoms form a honeycomb lattice and is sandwiched between two atomic
planes of I.  The Cr ions  are surrounded by 6 first neighbor I atoms arranged
in a corner sharing octahedra.  
In an ionic picture,
the oxidation state of Cr in this
compound is expected to be $+3$,  with  an electronic configuration $3s^0
3d^3$.  In an octahedral environment the $d$ levels split into a higher energy
$e_g$ doublet and a lower energy $t_{2g}$ triplet.\cite{abragam2012}
 Thus, we expect that Cr$^{3+}$ ions in this environment have $S=3/2$, with 3
electrons occupying the $t_{2g}$ manifold,  complying with first Hund rule (see Fig (\ref{fig1})c.
 The
lack of orbital degeneracy results in  an orbital singlet\cite{abragam2012}
with a  quenched orbital moment.  This
picture is consistent with the  observed\cite{mcguire2015} saturation
magnetization of bulk CrI$_3$, that yields  a magnetic moment of $\simeq$
3$\mu_B$ per Cr atom, that can be explained with $S=3/2$ and $L=0$. 

Single ion magnetic anisotropy is originated by the interplay of spin orbit
coupling and the crystal field. In  magnetic ions with  a finite orbital
moment,  magnetic anisotropy scales like  ${\cal E}_{\rm MAE}\propto \lambda
\langle \vec{L}\rangle\cdot \langle \vec{S}\rangle$, where $\lambda$ is the
magnetic ion atomic spin orbit coupling. However,  when the  orbital moment is
quenched ($\langle \vec{L}\rangle=0$),  this lowest order non-zero contribution
arises from quantum fluctuations of the orbital moment, and is given by
${\cal E}_{\rm MAE}\propto \frac{\lambda^2}{\Delta}$, where $\Delta$ is the
energy separation with the crystal field
excited states of the ion. Given that  
{$\lambda \simeq$ { 10} meV }
for Cr,\cite{NIST} and $\Delta$ is in the range of $500$
meV,  single ion anisotropy energies are very often way below 1 meV.    In a
purely octahedral environment this quadratic contribution would actually
vanish,\cite{abragam2012,ferron2015} and the magnetic anisotropy energy would scale like
${\cal E}_{\rm MAE}\propto \frac{\lambda^4}{\Delta^3}$, resulting in an
extremely small single ion anisotropy.   Based on these considerations,  single
ion anisotropy of Cr$^{3+}$ in CrI$_3$ should arise from the distortion of the
octahedral environment.

 Magnetic interactions between magnetic ions  separated by non-magnetic
ligands    arise via the super-exchange mechanism proposed by P. W.
Anderson.\cite{anderson1950} This involves the virtual excitation of excited
states where charge is transferred,  during a Heisenberg time, from the ligand
to the magnetic cations.   This virtual processes reduce the total energy of
the system and depend on the relative spin orientation of the magnetic atoms.
The sign of this exchange interaction depends both on the angle $\theta$ formed
by the two chemical bonds connecting the ligand and the magnetic atoms and on
the filling of the $d$ levels of the cations. A set of rules to predict the
sign of the interactions was proposed, independently,  by J. B.
Goodenough\cite{goodenough1958} and Kanamori.\cite{kanamori1959}  In particular,  ferromagnetic
interactions are maximal   when the  $\theta= 90^{o}$. For  CrI$_3$, the 
{angle
$\theta\simeq 93^o$}, which  accounts\cite{Feldkemper98} for the
ferromagnetic interactions. 
  As long as spin-orbit interactions are neglected, these exchange interactions
are always spin rotational invariant and can be described with a Heisenberg
coupling $J \vec{S}_1\cdot\vec{S}_2$. 
 
The possibility of magnetic anisotropy in the superexchange interactions in
magnetic insulators  was proposed early on by T. Moriya.\cite{moriya1960}  In
his seminar work, he considered the anisotropic interactions originated by
spin-orbit coupling in the magnetic ions. He found two types of  addition to
the Heisenberg coupling. The first   are  the Dzyaloshinski-Moriya (DM) term or
antisymmetric exchange,  
$\vec D_{ij}\cdot\left(\vec{S}_i\times\vec{S}_j\right)$, postulated
by Dzyaloshinski.\cite{dzyaloshinsky1958}
The second is the anisotropic symmetric exchange,  
$\lambda S^z_i S^z_j$.

In the case of exchange mediated by an anion, the DM
vector can be written as\cite{PhysRev.126.896}
$\vec D_{ij} = \vec r_i \times \vec r_j$,
where $\vec r_i, \vec r_j$ link the anion with the two magnetic atoms.
The DM favors non-collinear ground states.  However,
this term is absent in the CrI$_3$ crystal, since
the two paths mediated by iodine contribute to with a DM vector
with opposite sign that yield a net zero contribution.
In contrast, the anisotropic symmetric exchange term is allowed by symmetry
and, as we show below, it is definitely important in CrI$_3$.  The symmetric
and antisymmetric contributions to the anisotropic superexchange scale with
$\lambda_I^2$ and $\lambda$, respectively,\cite{moriya1960} where
$\lambda_I\simeq 0.6$eV, is the atomic spin orbit coupling of
iodine.\cite{NIST}

\begin{figure*}
 \centering
                \includegraphics[width=0.9\textwidth]{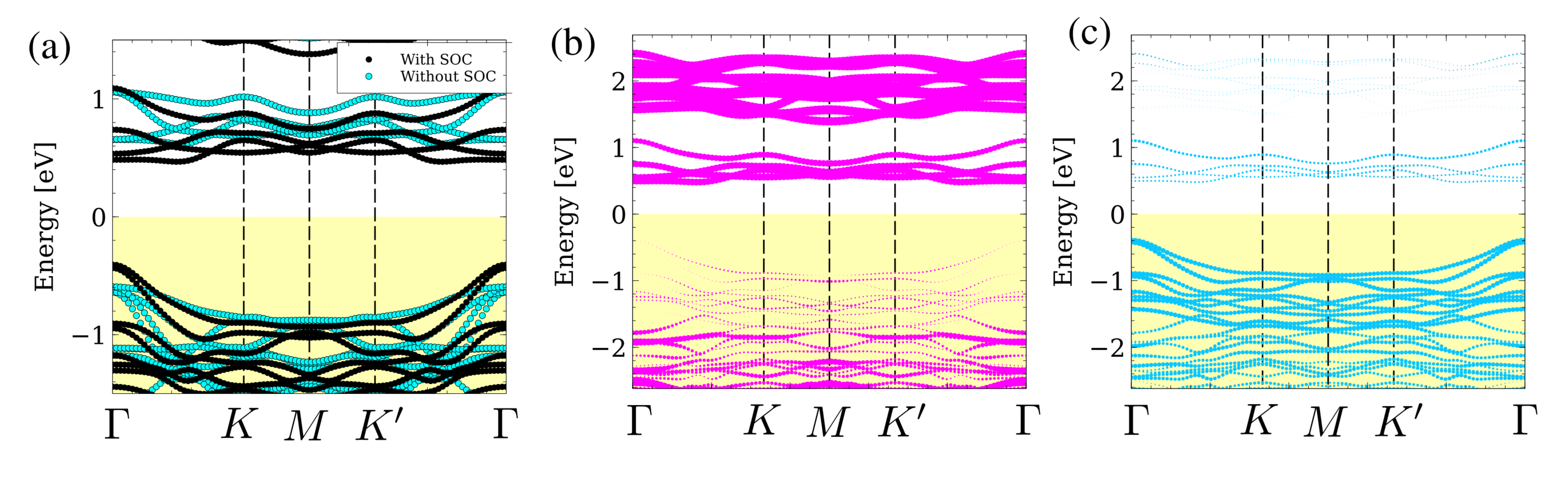}

\caption{
(a) Band structure of the system in the ferromagnetic state,
without SOC (light blue) and with SOC (black).
The large   variation upon switching on spin orbit coupling cannot be accounted by the energy scale of the spin orbit
coupling
of chromium.
Projection of the band structure over the chromium (b)
and iodine (c) atoms. From panels (b,c)
is inferred that the valence band shows a strong iodine character
and that the conduction band is dominant from chromium, but
with a non zero contribution from iodine. This chromium-iodine
mixing is the responsible of the large variation of the
conduction band upon switching on SOC, and is the
ultimate responsible of the anisotropic exchange.
}
\label{fig2}
\end{figure*}

\section{Density functional methods}
We perform density functional theory calculations with the 
pseudo-potential code Quantum Espresso\cite{QE-2009} and the all-electron code
Elk\cite{Elk}.
Monolayer structures were relaxed with Quantum Espresso, Projector augmented wave (PAW)
pseudopotentials\cite{PhysRevB.50.17953,kucukbenli2014projector}
and PBE exchange correlation functional\cite{PhysRevLett.77.3865} in the
ferromagnetic configuration. With the relaxed structures, calculation
with Elk  are carried out using spin orbit coupling in the
non-collinear formalism, DFT+U with the
Yukawa scheme\cite{PhysRevB.80.035121} ($J=0.7$ eV and $U=2.7$ eV)
in
the fully localized limit and LDA exchange correlation functional.\cite{PhysRevB.23.5048}
We  have verified that exchange energies with LDA or GGA, with
or without DFT+U 
give qualitatively similar results.

The calculations of magnetic anisotropy require careful
convergence of the total energy. 
We found that converging the total energy
{$10^{-8}$} eV yields stable results.
We have used the feature of Elk  that permits to tune the overall strength of
spin orbit interaction by a dimensionless constant scale factor, that we call
$\alpha$. Thus, for $\alpha>1$ the size of the spin orbit coupling is
increased above its actual value. In addition, we have introduced a
modification in the source code of Elk
in order to selectively  turn on and
off the spin orbit coupling in the two different atoms independently, so that
we now have two dimensionless scale factors, $\alpha_I$ and $\alpha_{Cr}$.  As
we discuss below, these two resources permit to 
to trace the origin of the magnetic anisotropy, as we discuss now.

\section{Electronic properties of CrI$_3$}
We now describe the  most salient electronic properties of CrI$_3$, as
described within our
DFT calculations, in line with previous work\cite{zhang2015,liu2016exfoliating}. 
The calculations show that CrI$_3$ is a ferromagnetic
semiconductor. The magnetic moment resides mostly in the Cr atoms, with a
residual counterpolarized magnetization on the $I$ atoms.  The total magnetic
moment in the unit cell is 6 $\mu_B$, {3}$\mu_B$  per Cr atom.
Figure \ref{fig2}a shows the band structure, calculated with and without SOC. 
The bands undergo a rather large 
shift,  in the range of $0.1$ eV, when SOC is included,.  The size
of this shift  is a first indication that the spin orbit interaction of iodine
atoms plays an important role,\cite{PhysRevB.86.134413} 
as spin orbit coupling in Cr is much smaller
than $0.1$ eV.
Figure \ref{fig2}b,c shows the bands weighted over the projection on the $d$
orbitals of Cr (Fig. \ref{fig2}b) 
and the $p$ orbitals of I (Fig. \ref{fig2}c).  It is apparent that  the
top of the
valence band is formed mostly by spin unpolarized $p$ orbitals of the I atoms,
and the conduction band is formed by $d$ orbitals of Cr.
The lowest lying states of the conduction band are majority $e_g$
orbitals, 
(around 0.7 eV in Fig. \ref{fig2})
whereas the minority states are located at higher energies
(around 2 eV in Fig. \ref{fig2}).
 The
majority spin $d$ orbitals, of the $t_{2g}$ manifold, are found $2$ eV below the
top of the valence bands.\footnote{In the case $U=0$, the 
weight of the $t_{2g}$ orbitals at the top of the valence band increases.}
The shape of the magnetization  field, not shown,  clearly shows that the
magnetic moment resides in orbitals with $t_{2g}$ symmetry, in line with
previous results.\cite{liu2016exfoliating}


\section{Magnetic anisotropy}
We are now in position to discuss the main topic of this work, magnetic
anisotropy. We have verified that the in-plane anisotropy is negligibly small. Therefore, in the following we
focus on the off-plane anisotropy and  we compute the quantity: 
\begin{equation}
{\cal E}_{\rm MAE}= E_G(0) - E_G(90)
\label{MAE0}
\end{equation}
where $E_G(\theta)$  is the computed ground state energy as a function of the
angle $\theta$ that forms the magnetic moment with the atomic planes.  ${\cal E}_{\rm MAE}>0$ 
describes an off-plane easy axis system.
  For the
in-plane component, we take $M_y=0$.    In line with
previous work,\cite{zhang2015}
we obtain
${\cal E}_{\rm MAE}=0.65$ meV.  Thus,  the calculation predicts that the
system has an easy axis anisotropy, perpendicular to the atomic planes, in
agreement with the experiments.\cite{huang2017}

\begin{figure}[t!]
 \centering
                \includegraphics[width=\columnwidth]{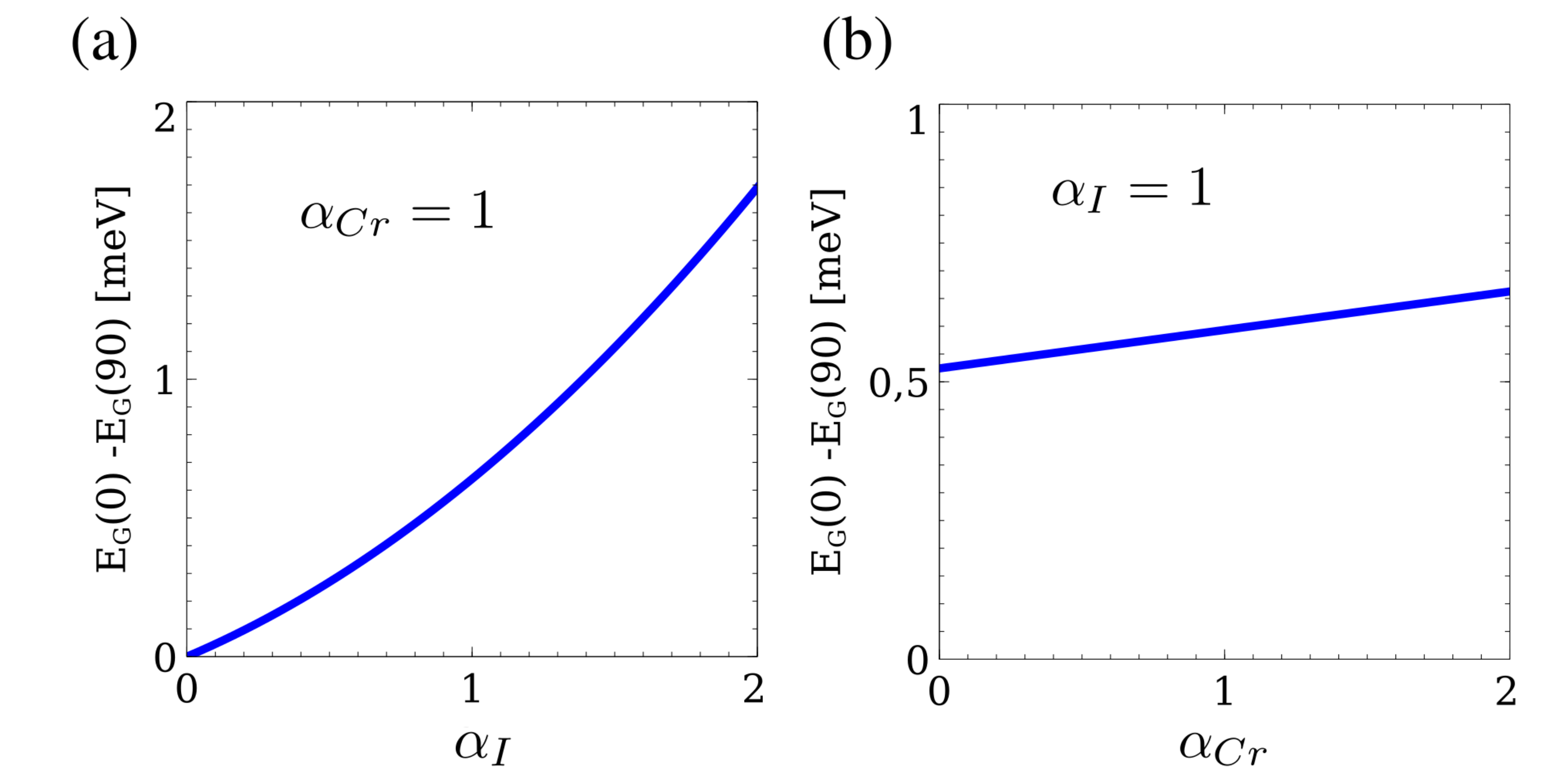}

\caption{(a) Evolution of the magnetic anisotropy energy as a function
of the spin-orbit coupling in iodine $\alpha_I$, keeping the
spin-orbit coupling in Cr to the real value $\alpha_{Cr}=1$
(b)  Same, reverting the roles of I and Cr.
These  curves show the  dominant contribution
of iodine spin orbit coupling to the MAE.
}
\label{fig3}
\end{figure}
In order to study the origin of this magnetic anisotropy we 
compute how ${\cal E}_{\rm MAE}$  changes
as we vary  independently  spin orbit coupling in two atoms.\cite{kosmider13}
To do so, here we define the DFT Hamiltonian as
\begin{equation}
{\cal H}_{\rm DFT} (\alpha_I,\alpha_{Cr}) = {\cal H}_0 + \alpha_I {\cal H}^{SOC}_I + 
\alpha_{Cr} {\cal H}^{SOC}_{Cr}
\end{equation}
where $\mathcal{H}_0$ is the non relativistic Hamiltonian,
$\mathcal{H}_{Cr}$ the relativistic Hamiltonian correction to chromium
and $\mathcal{H}_I$ the relativistic Hamiltonian correction to iodine.
We compute magnetic anisotropy energy from Eq. \ref{MAE0}, 
keeping at the default value $\alpha_{Cr,I}=1$ only one of the species, and
ramping the other.  The results are shown in Figs. 
\ref{fig3}a,b and permit to  conclude that MAE arises predominantly from the spin orbit
coupling in iodine atoms.  This suggests that anisotropic symmetric
superexchange is the likely cause of magnetic anisotropy in this compound. This
also seems to indicate that the local moments do not have a strong single ion
anisotropy, and therefore they are not properly described as Ising spins.

\subsection{Spin Hamiltonian}

  In order to validate these hypothesis, we now propose a model Hamiltonian for
the spins of the Cr atoms in the honeycomb lattice:
\begin{equation}
{\cal H}= -\left(\sum_i D (S^z_i )^2  + \frac{J}{2} \sum_{i,i'}\vec{S}_i
\cdot\vec{S}_{i'} + \frac{\lambda}{2} \sum_{i,i'} S^z_iS^z_{i'}\right)
\label{hamil}
\end{equation}
where the sum over $i$ runs over the entire lattice of Cr atoms, and the sum
over $i'$ runs over the 3 atoms, the first neighbors of atom $i$.   The first
term in the Hamiltonian describes the easy axis single ion anisotropy and we
choose $z$ as the off-plane direction.  The second term is the Heisenberg
isotropic exchange and the final term is the anisotropic symmetric exchange.
  The sign convention is such
that $J>0$  favors ferromagnetic interactions and $D>0$ favors off-plane easy
axis.  $\lambda=0$ would imply a completely  isotropic exchange interaction. 

We first treat Eq. \ref{hamil} in the classical approximation, and we 
describe the spins $\vec{S}$ as dimensionless classical vectors of length $S$ in the sphere
We write the energy of the ground state for 4 possible ground states, depicted
in Fig. \ref{fig4}a:  (I) ferromagnetic off-plane (FM,z) , (II)
antiferromagnetic off-plane (AF,z),  (III)
  ferromagnetic in-plane (FM,x) and (IV) antiferromagnetic in-plane (AF,x). We
denote the
  corresponding classical ground state energies as ${\cal E}_{FM,z},
 {\cal E}_{AF,z}, {\cal E}_{FM,x},{\cal E}_{AF,x}$.  
The spin model allows to write
the energetics of the different configurations
normalized per unit cell (2 Cr atoms) as 
\begin{eqnarray}
{\cal E}_{FM,z}= -2S^2D - 3S^2(J+\lambda) \\
{\cal E}_{AF,z}= -2S^2D +  3S^2(J+\lambda)  \\
{\cal E}_{FM,x}= -3S^2J  \\
{\cal E}_{AF,x}=+ 3S^2 J
\end{eqnarray}
with $S=3/2$ for CrI$_3$.
In order to determine  $J$, $D$ and $\lambda$, we use
the ground state energies for these 4 configurations as obtained from our DFT calculations.  In addition, we
do this ramping
the overall strength of the spin orbit coupling,
$\alpha=\alpha_{Cr}=\alpha_{I}$. 
For $\alpha=1$ we obtain $J= 2.2$ meV, in line with the results by
Zhang {\em et al.}\cite{zhang2015}
Our results  for $D$ and $\lambda$ are shown in Fig. \ref{fig4}b.  It is
apparent
that  the anisotropic symmetric exchange $\lambda$ is much bigger than the
single ion anisotropy $D$, in particular
for $\alpha=1$.  
The precise value of $D$ was affected by numerical noise in the regime
where both $J$ and
$\lambda$
already reached convergence, being always $D$
at least 30 times smaller that
the anisotropic exchange $\lambda$. This yields a value of D negligible with respect
any other exchange energy scale.
Thus, we have $J>\lambda
>> D$, which lead us to claim that the adequate spin model for CrI$_3$ is the
XXZ model with negligible single ion anisotropy.  This is the most important
result of this work. 
We find $\lambda= 0.04J = 0.09$ meV for $\alpha=1$. Thus, 
the flip-flop exchange is just 4 percent smaller than  the  Ising  exchange 
$S^z_i S^z_j$, given by $J+ \lambda$.  
Whereas the spin-flip part of exchange is actually responsible of
the existence of dispersive spin wave excitations, the anisotropic  term
$\lambda$ opens up a gap
in their spectrum, as we show below. This actually controls the
transition from the ferromagnetic to the non-magnetic phase as the material is
heated above $T_c$:

\begin{figure}
 \centering
                \includegraphics[width=\columnwidth]{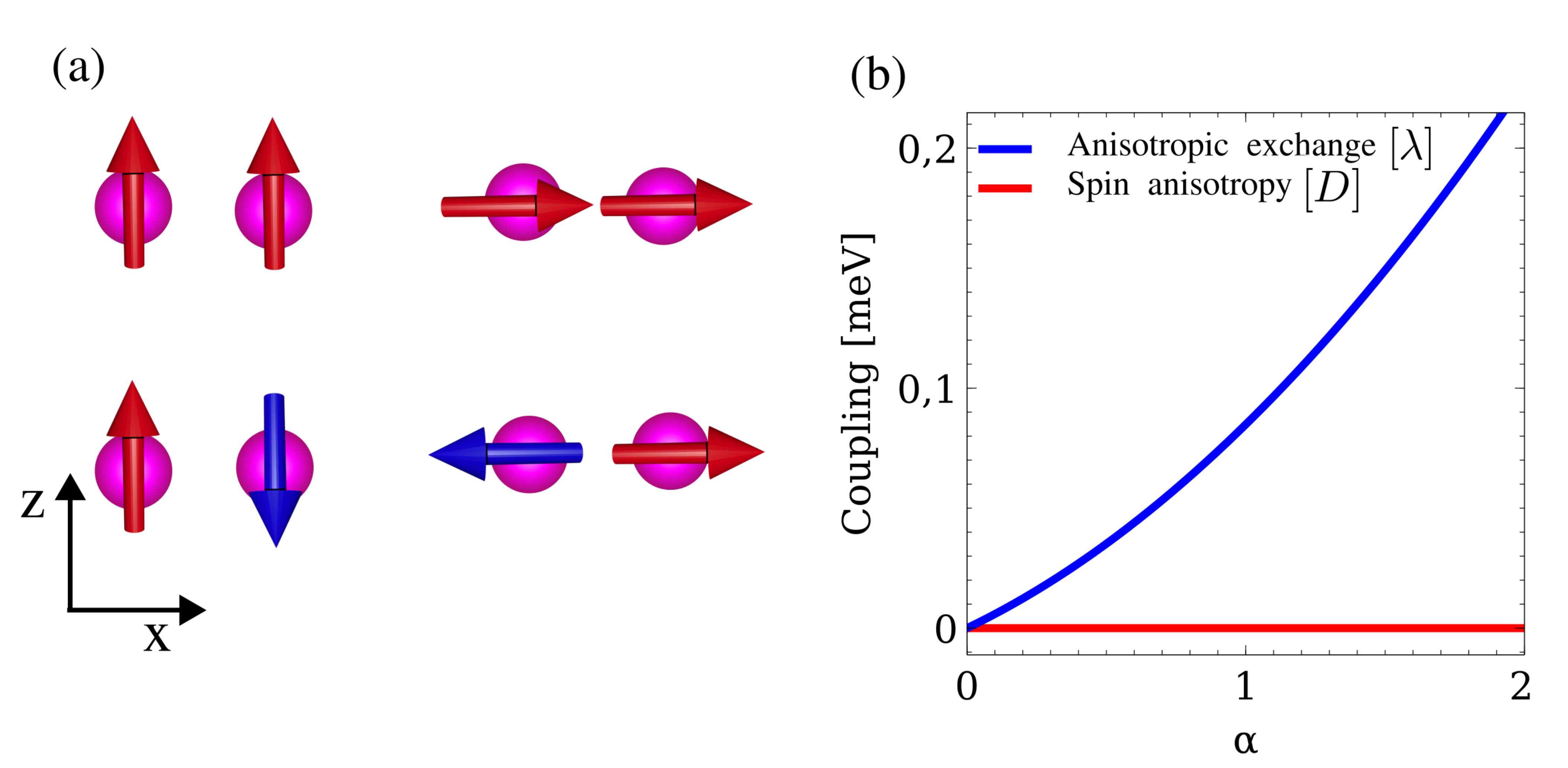}

\caption{
(a) Sketch and energetics of the different collinear magnetic configurations for
two chromium atoms, showing in-plane and off-plane ferromagnetic and
antiferromagnetic configurations.  Comparison with DFT permits
to extract $J$, $D$ and $\lambda$ (see text).
(b) Evolution of the single ion anisotropy $D$  and anisotropic exchange $\lambda$
as a function of the spin orbit coupling $\alpha$, where
  $\alpha=1$ corresponds to the  real value, as described at the DFT level.
}
\label{fig4}
\end{figure}

\section{Spin Wave theory}
We now go beyond the classical approximation used in the previous section. To do that, we now treat the spins in Hamiltonian (\ref{hamil}) as quantum mechanical $S=3/2$ operators. 
We treat the Hamiltonian within the linear spin wave approximation. To do so,
we use 
the so called
Holstein-Primakoff  representation
\cite{holstein1940field}
of the spin operators in terms of bosonic
operators  
\begin{eqnarray}
S^+_i = \sqrt{2S} \sqrt{1-\frac{b_i^\dagger b_i}{2S}} b_i \\
S^-_i = \sqrt{2S} b_i^\dagger \sqrt{1-\frac{b_i^\dagger b_i}{2S}}  \\
S^z_i = S - b_i^\dagger b_i 
\end{eqnarray}
with $b_i$ and $b^\dagger_i$ the bosonic annihilation and creation
operator in site.  The representation of the spin Hamiltonian (\ref{hamil}) in
terms of this bosonic operators leads a complicated non-linear Hamiltonian. The
spin wave approximation consist in keeping only the quadratic terms in the
bosonic operators $b$.  This approximation is valid for a small occupation of
the bosonic modes,  ie, when the magnetization is closed to $S_z\simeq S$, ie,
for small temperatures.  In the spin wave approximation, the effective
Hamiltonian for the spin excitations reads:
\begin{equation}
\mathcal{H}_{\rm spin \, waves} =
 \sum_i \left(2 D S+  3 S( J+ \lambda) \right)  b^\dagger_i b_i  
-JS  \sum_{\langle ij \rangle} b^\dagger_i b_j
\end{equation}
where the sum over $i$ runs over the entire lattice and the sum over $j$ runs
over the first neighbors of $j$. 
This Hamiltonian describes bosonic excitations moving in a honeycomb lattice,
with an on-site energy $\epsilon_0 = 2 DS +3S (J+\lambda)$ and a hopping energy
$JS$.  Thus, the Bloch Hamiltonian for the honeycomb lattice reads
\begin{eqnarray}
{\cal H}_{SW}(\vec{k}) = \left(\begin{array}{cc} 
\epsilon_0 & - JS f(\vec{k}) \\
-JS f^*(\vec{k}) &  \epsilon_0 
\end{array}
\right)
\end{eqnarray}
where $\epsilon_0 = 3JS + 2SD + 3S\lambda$,
$f(\vec{k})= 1  + e^{i\vec{k}\cdot\hat{a}_1}+e^{i\vec{k}\cdot\hat{a}_2}$
is the usual form factor for the honeycomb lattice, and $\hat{a}_{1,2}$ are the
unit vectors of the triangular lattice. The resulting energy spectrum is 
\begin{equation}
E^{\pm}(\vec k)  = \epsilon_0 \pm JS  \sqrt{|f(\vec{k}|^2}
\end{equation}
We can expand the lower band around its minima at the $\Gamma$ point, to get
\begin{equation}
E^{-}(\vec k)\simeq \Delta_0 + \rho k^2
\label{low}
\end{equation}
where the spin wave gap is given by
\begin{equation}
\Delta_0= 2DS+ 3 S \lambda
\label{SWG}
\end{equation}
For CrI$_3$ we can take $D=0$ and  we have a spin wave gap  $\Delta_0=3S\lambda=0.4$ meV.
The so called spin stiffness is given by
\begin{equation}
\rho = \frac{1}{4}JS
\end{equation}
that yields for CrI$_3$ a value $\rho = 0.82$ meV.  The ratio
 $\frac{\Delta_0}{\rho}=\frac{12 \lambda}{J}\simeq0.49$ plays an
important role in the following.

\begin{figure*}
 \centering
                \includegraphics[width=.9\textwidth]{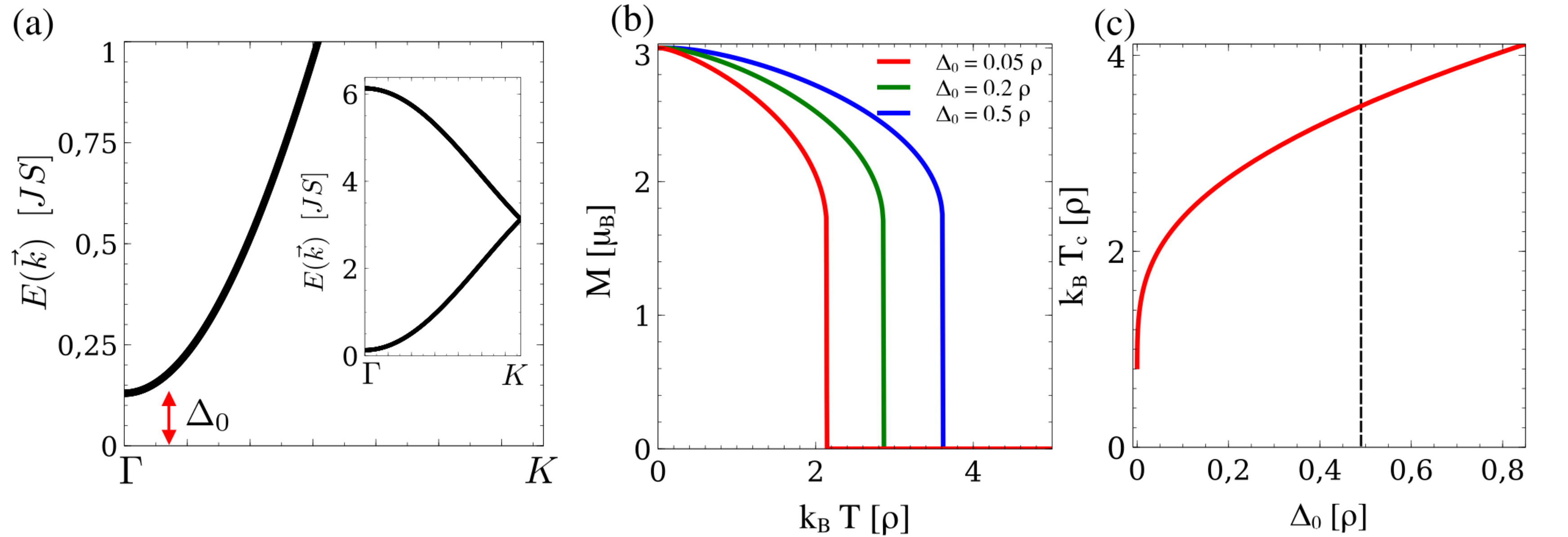}

\caption{(a) Example of the spin wave dispersion, showing
a gap in the spin excitations. (b) Selfconsistent solution of the
magnetization derived with Eq. \ref{scfm}, showing a 
depleted magnetization with increasing temperature.
(c) Numerical solution for the critical
temperature as a function of the spin wave gap $\Delta_0$,
showing a logarithmic dependence on $\Delta_0$. 
The dashed line in (c) is the result obtained for CrI$_3$
with the DFT calculations.
}
\label{fig5}
\end{figure*}

From Eqs. \ref{low},\ref{SWG} it is apparent that if the two terms that break spin rotational invariance in
the original Hamiltonian (\ref{hamil}),  $D$ and $\lambda$, vanish, the spin
wave spectrum becomes gapless. Therefore, in the spin wave spectrum,
both the anisotropic exchange and the single ion anisotropy
create a gap in the spin waves 
(see Fig. \ref{fig5}a), so that their effect on the
spin wave dispersion is similar.
This implies that simple inspection
of the spin wave dispersion does not provide enough information
to asses whether if the correct model for a compound is 
single ion anisotropy or anisotropic exchange,
and input from a microscopic first principles calculation
is necessary.
As we discuss now, the presence of their induced gap is
essential to have magnetization at finite temperature.

\subsection{Low temperature magnetization}
Every magnon carries one unit of angular momentum. Therefore, in the  linear
spin wave framework, we can approximate  the  magnetization by
\begin{equation}
M(T)= S -\delta M =  S - \frac{1}{2(2\pi)^2} 
\int_{BZ} \frac{d^2\vec k}{e^{\beta E(\vec k)}-1}
\label{MT}
\end{equation}
where $M(T)$ is the magnetization
in units of $\hbar$ 
per Cr atom as a function
of the temperature, the
factor $1/2$ comes from having 
two Cr atoms in the unit cell and $\beta^{-1}=k_BT$. 
Linear spin wave theory works well when
the magnon density  is small. 
In the following we use the fact that this
integral is controlled by the low energy magnons, and we approximate
$E(\vec k)$ by
Eq. (\ref{low}). In addition, we replace the integral over the Brillouin zone
by the integral over a circle of radius $k_c$, chosen
so that the density of magnons is properly normalized
$\frac{1}{2\pi}\int_0^{k_c} k dk = 2$.
Choosing this normalization includes
the contribution from the high energy magnon
branch in the high energy part of the dispersion.
We first focus in the case $\Delta_0=0$, i.e. in the absence of
anisotropic exchange, in that case the correction of the magnetization goes as

\begin{equation}
\delta M=  \frac{1}{4\pi} 
\int^{k_c}_{0} \frac{k dk}{\beta \rho k^2} \rightarrow \infty
\end{equation}
This  divergence signals the absence of order at finite
temperature in the Heisenberg model
in the gapless regime $\Delta_0=0$,
consistent with the Mermin-Wagner theorem.\cite{mermin66}
Therefore,  the anisotropy gap is essential to protect the long range order in
2D.  

We will move now to the case of finite spin wave gap $\Delta_0\ne 0$.
We now consider the very low temperature case, $\beta \Delta_0>>1$.  We can
then approximate\cite{PhysRevB.43.6015}
\begin{equation}
M(T) = M(T=0) + \delta M =  S -  \frac{ k_B T}{2\pi JS }e^{-\Delta_0/k_B T} 
\end{equation}
Thus, we expect that the magnetization will have a very weak temperature
dependence for temperatures smaller than  spin wave gap. According to  our
calculations  $\Delta_0= 0.4$ meV, so $M(T)$ 
be almost maximal up
to $T=5$ K.

\subsection{Estimate of $T_c$}
We now provide a rough estimate of the Curie temperature, based on {\em non linear}
spin wave theory . We use the initial expression for spin operators, and
expand them retaining the up to fourth order in the bosonic operators
\begin{eqnarray}
S^+_i \approx \sqrt{2S} \left (1-\frac{b_i^\dagger b_i}{4S}\right ) b_i \\
S^-_i \approx \sqrt{2S} b_i^\dagger \left (1-\frac{b_i^\dagger b_i}{4S} \right )  \\
S^z_i = S - b_i^\dagger b_i 
\end{eqnarray}

At intermediate temperatures, there is a finite number
of spin waves, that is accounted by the higher order terms
in bosonic operators when substituting the previous expansion
in the spin Hamiltonian.
In that situation, the spin Hamiltonian contains
four field operators and therefore is not exactly solvable.
Thus,  the effect of the spin wave population
is described using a mean field approximation 
in the spin wave Hamiltonian
by means of the substitution
$b_i^\dagger b_i b_j^\dagger b_j \approx 
\langle b_i^\dagger b_i \rangle b_j^\dagger b_j +
b_i^\dagger b_i \langle b_j^\dagger b_j \rangle + \mathcal{C}$.
With the previous approximation it is straightforward to check
that a finite population of spin waves is equivalent to a renormalization
of the hopping energy and spin wave gap as\cite{PhysRevB.84.064505}

\begin{equation}
JS \rightarrow J(S - \langle b^\dagger b \rangle) = JM(T)
\end{equation}

\begin{equation}
\lambda S \rightarrow \lambda (S - \langle b^\dagger b \rangle) =\lambda M(T)
\end{equation}
The previous substitutions lead to a selfconsistent equation for
the magnetization as

\begin{equation}
M = S - 
\frac{1}{2(2\pi)^2}
\int_{BZ} \frac{d^2 \vec k}{e^{\beta M E(\vec k)/S}-1}
\label{scfm}
\end{equation}
where the integral extends over the first Brillouin zone.
A qualitative behavior of the previous integral can be
obtained approximating $E(\vec k) = \Delta_0 + \rho k^2$
and 
$e^{\beta \frac{M}{S}E(k)}-1  \approx  \beta M (\Delta_0+\rho k^2)/S$.
As Eq. (\ref{scfm}) has no solution for $M=0$,  we define 
$T_c$
 as the temperature at which the magnetization is
depleted to $M=S/2$. This leads to the following equation:
\begin{equation}
k_B T_c \simeq  \frac{2\pi \rho S}{ \log \frac{\Delta_0 + 8\pi\rho}{\Delta_0}}
=\frac{\pi JS^2 }{2  \log \frac{\Delta_0 + 2 \pi JS}{\Delta_0}}
\label{tc}
\end{equation}
A very similar result can be  obtained  using different spin representations.\cite{irkhin1999,grechnev2005}
Equation (\ref{tc}), together with the
numerical solution\cite{PhysRevLett.9.286} of Eq. (\ref{scfm})
in Fig(\ref{fig5})b, 
show several 
important results. 
First,  $T_c$  is an increasing function of the spin wave gap $\Delta_0$ (see Fig. (\ref{fig5}c).
This is in line with the experimental results recently reported for Cr$_2$Ge$_2$Te$_6$,\cite{gong2017}
for which the major contribution to the spin wave gap comes from the Zeeman contribution, due to
the  very tiny intrinsic anisotropy, resulting in dramatic variations of $T_c$
as a function of the applied field.  This is a feature specific of two
dimensional magnets with dispersive spin waves. 
Second, $T_c$ is significantly smaller than the prediction coming from the
Ising model. The exact solution for the Ising model in the honeycomb
lattice\cite{meyer2000} yields $k_BT_c= 1.51 j$, where $j$ is the coupling
between classical spins with $S=1$.   Using  this result for CrI$_3$, we would
have $k_BT_c= 1.51 (J+\lambda) S^2= 85$ Kelvin, that overshoots the
experimental value 45 K. 

On the other hand,  
using the prediction of $T_c$ obtained by the numerical solution of 
Eq. (\ref{scfm})
shown in Fig. \ref{fig5}c, we obtain a value of $k_BT_c= 3.5\rho $,
for $\Delta_0 =0.49 \rho$,
that gives $T_c= 33$ K, 
understimating the
the experimental value\cite{huang2017}
$T_c =45$ K by 20\%.
Including the effect of the finite magnetic field
would increase $\Delta_0$, and push
the prediction upward.   
Inclusion of longer range coupling\cite{PhysRevB.91.235425,
PhysRevB.83.125126,PhysRevB.88.075106,PhysRevB.88.035107}
is also expected to
increase the spin stiffness, yielding a larger estimate
of the critical temperature.
Furthermore, a more accurate treatment must consider the explicit spin wave
density of states and a more careful treatment of fluctuations
close to the critical point.
The discrepancy highlights the limitations of the non-linear spin wave theory,
and perhaps, also those of the DFT scheme to determine the energy scales
of the Hamiltonian.
Nevertheless, apart from the previous limitations,
our approach highlights the role played by
anisotropic exchange, as the ultimate mechanism responsible to controlling
the divergence in Eq. \ref{tc}.

\section{Conclusions}
We have studied the origin of magnetic anisotropy in two dimensional CrI$_3$, a recently
discovered ferromagnetic two  dimensional crystal with off-plane anisotropy.
We have found that  magnetic anisotropy in this system
 comes predominantly from
the superexchange
interaction, that gives rise to an anisotropic
contribution to the conventional exchange interaction.
The strength of the non Heisenberg correction is found to be controlled
by the spin orbit coupling of the
intermediate iodine atom. The single ion anisotropy of the magnetic
Cr atoms
is found to give a negligible contribution to magnetic anisotropy.
The
suppression of the single ion anisotropy
due to the octahedral environment, together
with large spin orbit coupling of iodine,
make  the anisotropic exchange the leading mechanism
stabilizing the magnetic ordering in 2D CrI$_3$.
Our calculations permit to conclude that the
effective spin Hamiltonian for CrI$_3$ is a XXZ model. In turn, this implies that
gapped spin waves are the essential elementary excitations that control
the finite temperature properties of this new  type of magnetic system.  Given that spin waves in two dimensions are  interesting on its own right, as they can exhibit thermal Hall effect and have
topologically non-trivial
phases.\cite{hirshberger2015,chisnell2015,roldan2016,owerre2016,zyuzin2016,PhysRevLett.117.217202}
{
As an example, one can consider inducing a Dzyaloshinskii-Moriya
term in a
CrI$_3$ monolayer by applying a perpendicular electric field,
opening the possibility of a skyrmionic ground state whose magnonic
Hamiltonian is topologically non-trivial and shows gapless edge magnonic
excitations.\cite{roldan2016}
Another interesting playground would be the possibility of applying
non uniform strain to the ferromagnetic monolayer, modulating
the exchange constants and creating an artificial gauge field
in the
magnonic Hamiltonian.\cite{kraus2016quasiperiodicity,PhysRevLett.111.226401}
}
Therefore, the discovery of magnetic 2D crystals paves the
way towards the exploration of
these exciting phenomena.

\

\
\

\section*{Acknowledgments}
We acknowledge F. Rivadulla for fruitful conversations
and D. Xiao for useful remarks.
We acknowledge
financial support by Marie-Curie-ITN 607904-SPINOGRAPH. 
JFR acknowledges  financial supported by MEC-Spain (MAT2016-78625-C2).
 This work has been supported in part 
 by ERDF funds
through the Portuguese Operational Program for Competitiveness
and Internationalization COMPETE 2020,
and National Funds through FCT- The Portuguese Foundation
for Science and Technology, under the project
PTDC/FIS-NAN/4662/2014 (016656).
J. L. Lado thanks
the hospitality of the Departamento de Fisica Aplicada
at the Universidad de Alicante.

\bibliographystyle{apsrev4-1}
\bibliography{biblio}{}

\end{document}